\documentclass[11pt,a4paper]{article}
\pdfoutput=1 

\RequirePackage{ifpdf}
\usepackage{amsmath} 

\usepackage{mpaper}
\usepackage{pstricks}
\usepackage[final]{pdfpages}
\usepackage{ifpdf}
\usepackage{slashed}

\usepackage{hyperref}

\usepackage{color}
\usepackage{graphics}

\usepackage{etoolbox} 

\title{\sffamily Associated production of Higgs boson with vector boson 
at threshold N$^3$LO in QCD}

\author[a]{M.~C.~Kumar}
\author[b]{M.~K.~Mandal}
\author[a]{V.~Ravindran}

\affiliation[a]{The Institute of Mathematical Sciences, Chennai, India }
\affiliation[b]{Regional Centre for Accelerator-based Particle Physics,\\ Harish-Chandra Research Institute, Allahabad, India}

\emailAdd{mckumar@imsc.res.in}
\emailAdd{mandal@hri.res.in}
\emailAdd{ravindra@imsc.res.in}

\preprint{HRI-RECAPP-2014-027}
\abstract{We present the results for the associated production of Higgs boson with vector boson computed at threshold N$^3$LO in QCD. We use the recently available result on the threshold contributions to the inclusive Drell-Yan production cross-section at third order in strong coupling constant.  We have implemented it in the publicly available computer package {\tt vh@nnlo}, thereby obtaining the numerical impact of threshold N$^3$LO contributions for the first time.  We find that the inclusion of such corrections do reduce theoretical uncertainties resulting from the renormalization scale.}

\begin{document}
\allowdisplaybreaks[4]
\unitlength1cm
\maketitle
\flushbottom

\section{Introduction}
\label{intro}
Recently, LHC has discovered a particle with a mass at around 125 GeV~\cite{1207.7214, 
1207.7235}, whose production cross section and decay rates are compatible with those 
predicted for the Higgs boson of the Standard Model (SM). However, still we have to determine its quantum numbers (spin and CP 
properties), mass and the strength and nature of the couplings to other standard model 
particles accurately. The close inspection of the accumulating luminosity over a longer 
period will not only reveal the complete picture of the electroweak symmetry breaking sector 
but also any hint of new physics buried within the results.
Any discrepancies in its measured cross section from QCD calculations may signal deviations of the SM predictions. It is thus important to provide a precise
calculation of the production of the Higgs boson, 
with a reliable estimate of the theoretical error due to the missing higher order terms.
In the SM, the Higgs boson is produced mainly through gluon fusion, whereas the alternative channels 
are associated production with vector bosons namely $W/Z$ bosons, often called
Higgs-Strahlung process, vector boson fusion processes, bottom quark annihilation etc.
The Higgs bosons are produced primarily at the LHC via gluon gluon fusion through a top
quark loop~\cite{Georgi:1977gs, Djouadi:1991tka, Dawson:1990zj, Spira:1995rr, 
Catani:2001ic, Harlander:2001is, Harlander:2002wh, Anastasiou:2002yz, 
Ravindran:2003um}, which has been known up to next-to-next-to leading order 
(NNLO) in the literature for a long time. The sub-dominant channels for the 
production comprising of the vector boson fusion \cite{Han:1992hr, 
Bolzoni:2010xr} and associated production with vector bosons~\cite{Han:1991ia, 
Brein:2003wg} are also known up to NNLO accuracy in QCD.
The bottom-antibottom ($b \bar b$) annihilated inclusive production for the 
Higgs boson is also available at NNLO accuracy considering five active flavours 
i.e. including the bottom quarks in the parton density 
function~\cite{Dicus:1988cx, Dicus:1998hs, Maltoni:2003pn, Olness:1987ep, 
Gunion:1986pe, Harlander:2003ai}.   In the case of Higgs production, the scale 
dependence even at NNLO are not convincingly negligible due to potential missing 
higher terms.  Hence there is a constant pursuit of increasing the accuracy of 
the results with the systematic inclusion of higher order terms in QCD and there 
are on going efforts to go beyond the existing NNLO level. 
The partial results on inclusive Higgs boson production and also DY production 
beyond NNLO~\cite{Moch:2005ky, Laenen:2005uz, Idilbi:2005ni, Ravindran:2005vv, 
Ravindran:2006cg} have been reported after taking 
into account the dominant effects of the soft gluon radiations from higher 
orders, using the available results of the quark and gluon form factors 
\cite{Moch:2005id, Moch:2005tm, Gehrmann:2005pd, Baikov:2009bg, 
Gehrmann:2010ue}, the mass factorization kernels \cite{Moch:2004pa}, the 
renormalization constant \cite{Chetyrkin:1997un} for the effective operator 
describing the coupling between the Higgs boson and the SM fields in the 
infinite top quark mass limit and the NNLO soft contributions 
\cite{deFlorian:2012za} in $d$ dimensions. The reported threshold corrections, 
which manifest them through the delta function and the plus distribution of the 
logarithms, was partial in the sense that the complete distribution in 
connection with the delta function was not available. Since then, several 
advances have happened \cite{Kilgore:2013gba, Anastasiou:2013mca, Duhr:2014nda, 
Dulat:2014mda} with the aim of obtaining the complete next-to-next-to-next-to 
leading order (N$^3$LO) result for the inclusive rate of the Higgs boson 
production.
Recently, Anastasiou et al.~\cite{Anastasiou:2014vaa} have obtained the delta 
function part of the threshold N$^3$LO contribution for the inclusive Higgs 
boson production through gluon fusion. This generated a plethora of results at 
threshold N$^3$LO in QCD. These include inclusive DY 
production~\cite{Ahmed:2014cla,Li:2014bfa}, 
inclusive production of Higgs boson in bottom quark 
annihilation~\cite{Ahmed:2014cha} and the general expression of the hard-virtual 
coefficient~\cite{Catani:2014uta} combined with the threshold resummation at 
next-to-next-to-next-to-leading-logarithmic (N$^3$LL) accuracy for the 
production cross section of a colourless heavy particle at hadron colliders at 
threshold N$^3$LO. Apart from that, the rapidity distributions of the Higgs 
boson in the gluon fusion and bottom quark annihilation as well as the dileptons 
in DY  have been reported to this accuracy in the threshold limit, see ~\cite{Ahmed:2014uya, Ahmed:2014era}. 
In addition, progress in obtaining beyond the threshold corrections \cite{Presti:2014lqa, 
deFlorian:2014vta} for the inclusive Higgs production at N$^3$LO is already 
underway.  Recently, the full next to soft as well as the exact results for the 
coefficients of the first three leading logarithms at this order have been 
obtained for the first time in~\cite{Anastasiou:2014lda}.

The Higgs-Strahlung process is one of the potential channels for the Higgs boson 
production at the LHC. The LO amplitude is an electroweak process and hence, the 
higher order QCD corrections enter only in the initial state comprising of a 
quark and an antiquark. This fact prompted this process to be represented in 
terms of the convolution of production of a virtual $W$ or $Z$ boson production 
(DY like) and decay rate of that virtual boson to a real vector boson and the 
Higgs boson, at every order in QCD.
Therefore, the available higher order QCD corrections of DY like processes can 
be used to study the QCD effects in Higgs-Strahlung process.
The QCD corrections to DY at next-to leading order (NLO) \cite{Altarelli:1978id} 
as well as at NNLO~\cite{Hamberg:1990np, Harlander:2002wh} are known for long 
time and they have been already used in the Higgs-Strahlung process 
\cite{Han:1991ia, Baer:1992vx, Ohnemus:1992bd, Mrenna:1997wp, Brein:2011vx, 
Kniehl:1990iva, Brein:2003wg, Ferrera:2014lca, Ferrera:2011bk}.
At NNLO, for the associated production of the Higgs boson with $Z$ boson, there 
are additional corrections coming from the gluon fusion via a box diagram and 
also the quark antiquark initiated processes, where the Higgs boson is coupled 
to a top quark loop. These corrections have been obtained in 
~\cite{Kniehl:1990iva, Brein:2003wg, Brein:2011vx}. The gluon induced production of the associated Higgs boson has also been 
reported at NLO~\cite{Altenkamp:2012sx} and the threshold resummation has been 
completed at NLL 
accuracy~\cite{Harlander:2014wda}. There are no such additional 
corrections in the case of $W$ boson production with the Higgs boson as the 
final state not being a charge neutral. The {\tt vh@nnlo}~\cite{Brein:2012ne} 
program includes all these contributions separately for the $Z$ and $W$ boson 
production with the Higgs boson up to  NNLO. The electroweak (EW) corrections 
reported in~\cite{Ciccolini:2003jy, Brein:2004ue}, have been also incorporated 
in this programme as a multiplicative factor based on the fact that the EW 
corrections for these processes do not depend on any of the QCD parameters.  
The NLO corrections have been found to enhance the total inclusive rate  by 
31$\%$ whereas the NNLO DY terms contribute towards additional 3$\%$ correction 
for the $ZH$ production at LHC8. The numerical values for the $WH$ production 
are also very similar for the DY type corrections up to NNLO.
The top loop effect can be counted for 1$\%$ correction for both the processes. The additional gluon initiated box diagrams generate 5$\%$ correction in case of the $ZH$ production. 
These numerical values exhibit that the corrections at NNLO are small in size although it has been observed that scale dependence is reduced significantly at this level. However, the inclusion of higher order terms is important to assess the reliability of the purturbative calculations as well as to have a better understanding of the pattern of these corrections at higher orders.

The paper is organized as follows. In the Sec.~\ref{N3LOSV} we present the results contributing at N$^3$LO in threshold limit.   We then discuss the numerical impacts at LHC in Sec.~\ref{Results}. Finally, we conclude with our findings in Sec.~\ref{Conclude}. 

\section{Threshold Corrections Beyond NNLO}
The inclusive production of Higgs boson in association with vector boson come from factorizable and
non factorizable partonic subprocess.  The factorizable ones can be written as convolution of the production of virtual vector boson and its decay to Higgs boson.  They are often called DY type.   The DY type contribution to the hadronic cross-section $P \left(p_1\right)+P\left(p_2\right) \rightarrow V\left(p_V\right) + H\left(p_H\right)$ can be expressed as
\begin{equation}
\sigma \left(S, M_V^{2}, M_H^{2}\right) = \int_{(M_H+M_V)^{2}}^{S} {\rm d}q^{2}\sigma^{V^{*}} \left(q^{2},S \right) \frac{{\rm d}\Gamma \left(M_V^{2}, M_H^{2}, q^{2}\right)}{{\rm d}q^{2}}
\end{equation}
where, $p_1$ and $p_2$ are the incoming hadronic momenta and $S$ is the hadronic center of mass energy squared $\left(S \equiv \left(p_1+p_2\right)^2 \right) $. The corresponding one for the incoming partons at the partonic level is given as $\hat{s} = (k_1+k_2)^2$.  The momentum of the virtual gauge boson $V^{*}$ is $q=(p_V+p_H)$.
The parton level cross section for the production of virtual vector boson $V^*$ which is of DY type is denoted by $\sigma^{V^{*}}$ and $\frac{d\Gamma}{dq^{2}}$ is the decay rate of that virtual boson to a real vector boson and the Higgs boson and is given by:
\begin{equation}
\frac{ {\rm d} \Gamma \left(M_V^{2}, M_H^{2}, q^{2}\right)}{ {\rm d}q^2 } = \frac{ G_F M_V^4}{
2\sqrt{2} \pi^2}  \frac{\lambda^{1/2} (M_V^2, M_H^2; q^2)}{(q^2-M_V^2)^2}
\left(1 + \frac{\lambda(M_V^2, M_H^2;q^2)}{12M_V^2/q^2} \right)
\end{equation}
with, $\lambda\left(x,y;z\right)$ = $\left(1-\frac{x}{z}- \frac{y}{z}\right) ^2-4\frac{xy}{z^2}$, being the usual phase-space function for the two body final state. Now, the DY type production cross-section can be expressed as :
\begin{equation}
\sigma ^{V^{*}}(q^{2},S) = \frac{1}{S} \sum_{a,b} \int_{0}^{1}dx_{1} \int_{0}^{1}dx_{2}  \int_{\tau}^{1}dz
f_a\left(x_{1}, \mu_{F}^{2}\right) f_b\left(x_{2}, \mu_{F}^{2}\right) {\Delta}_{ab}^{V^{*}}
\left(z, q^{2}, \mu_{F}^{2} \right) \delta\left(\tau - x_{1}x_{2}z\right) 
\end{equation}
The $f_a$ and $f_b$ are the parton density function renormalized at $\mu_{F}$. We have defined ${\Delta}_{ab} \equiv \hat{s} \hat{\sigma}$ with $\tau = q^{2}/S $ and $z = q^{2}/\hat{s} $. This finite ${\Delta}_{ab}$ can be expanded in terms of the strong coupling constant as follows:
\begin{equation}
{\Delta}_{ab}\left(z, q^{2}, \mu_{F}^{2} \right) = \sum_{i=0}^{\infty} 
\left(a_s\left(\mu_{R}^{2}\right)\right)^i {\Delta}_{ab}^{(i)}\left(z, q^{2}, \mu_{F}^{2}, \mu_{R}^{2} \right)\, ,
\end{equation}
where, $a_s\left(\mu_{R}^{2}\right) = \frac{g_s\left(\mu_{R}^{2}\right)^2}{16 \pi^2}$.

Beyond LO (i.e. $i=0$) the purturbative coefficients $\Delta_{ab}^{(i)}$ can be split into two parts.
\begin{equation}
\Delta_{ab}^{(i)} (z,q^2,\mu_F^2,\mu_R^2) = 
     \Delta^{{\rm hard}, (i)}_{ab}(z,q^2,\mu_F^2,\mu_R^2)
   + \delta_{a q} \delta_{b \overline q} \Delta_{ab}^{\rm SV, (i)}(z,q^2,\mu_F^2,\mu_R^2)\,
\end{equation}
The hard part $\Delta^{{\rm hard}, (i)}_{ab}$ contains the regular terms in the variable $z$ and the SV part $\Delta^{{\rm SV}, (i)}$ is simply proportional to $\delta(1-z)$ and ${\cal D}_k$ resulting from virtual and soft gluon radiations, that is
\begin{equation}
\Delta_{ab}^{\rm SV, (i)}\left(z\right)=\Delta_{ab}^{\rm SV, (i), \delta}\delta\left(1-z\right)
+\sum_{k=0}^\infty \Delta_{ab}^{\rm SV, (i), (k)}{\cal D}_{k}
\end{equation}
with
\begin{equation}
{\cal D}_{k} = \left( ln^{k}(1-z) \over (1-z)\right)_{+}
\end{equation}
As we have already discussed, the hard and soft parts of $\Delta^{(i)}_{ab}$ are known up to
NNLO level in QCD.  At N$^3$LO level, only $\Delta^{\rm SV,(3)}_{q \overline q}$ 
is known, see \cite{Ahmed:2014cla,Li:2014bfa,Catani:2014uta}.
The computation of SV part of $\Delta^{(3)}_{q \overline q}$ in \cite{Ahmed:2014cla} uses
the factorization property of the QCD amplitudes and the Sudakov resummation of soft gluons.
At N$^3$LO level in QCD, SV part requires quark form factor as well as the diagonal
terms of the mass factorization kernels up to three loop level and the contributions 
of soft gluon radiations in the single, double and triple gluon emission 
subprocesses to third order 
in strong coupling constant.  While form factor
and the kernels are available to desired accuracy for quite some time, 
the third order soft gluon effects from real
emission subprocesses have been missing to
get N$^3$LO results till recently.  A spectacular achievement by Anastasiou et al. \cite{Anastasiou:2014vaa}
in obtaining the third order soft gluon radiations in the inclusive Higgs production 
and better understanding of the soft gluon resummation paved  the way to obtain several third order
results as has been discussed earlier in the Sec.~\ref{intro}.
%
Along this direction,
the results of \cite{Ahmed:2014cla,Li:2014bfa,Catani:2014uta} can be used in Higgs-Strahlung 
processes to get an estimate of the effects from threshold N$^3$LO DY type of corrections as
the threshold effects in DY production are found to be significant.
Up to NNLO, DY type of corrections can be found in \cite{Hamberg:1990np,Harlander:2002wh} and the 
threshold N$^3$LO DY correction with $\mu_R=\mu_F= Q$ is given here for completeness:
\label{N3LOSV}
\begin{align}
\Delta^{\rm SV, (3)}_{q \overline{q}} &= \delta(1-z) \Bigg(
C_A^2 {C_F} \Bigg(\frac{13264}{315} \;{\zeta_2}^3 + \frac{14611 \
}{135} \;{\zeta_2}^2 - \frac{884}{3} \;{\zeta_2} {\zeta_3} + 843 \
{\zeta_2} - \frac{400}{3} \;{\zeta_3}^2
\nonumber \\
&+\frac{82385}{81} \;{\zeta_3} 
- 204 \;{\zeta_5} 
- \frac{1505881}{972}\Bigg) 
+ {C_A} C_F^2 \
\Bigg(-\frac{20816}{315} \;{\zeta_2}^3 -\frac{1664}{135} \;{\zeta_2}^2 
+\frac{28736}{9} \;{\zeta_2} {\zeta_3}
\nonumber \\
&- \frac{13186}{27} \;{\zeta_2} 
+\frac{3280}{3} \;{\zeta_3}^2 
- \frac{20156}{9} \;{\zeta_3} - \frac{39304}{9} \;{\zeta_5}
+ \frac{74321}{36}\Bigg) 
 + {C_A} {C_F} {n_f} \
\Bigg(-\frac{5756}{135} \;{\zeta_2}^2
\nonumber \\
&+ \frac{208}{3} \;{\zeta_2} {\zeta_3}
- \frac{28132}{81} \;{\zeta_2}
- \frac{6016}{81} \;{\zeta_3} - 8 \;{\zeta_5} + \frac{110651}{243}\Bigg)
+ C_F^3 \
\Bigg(-\frac{184736}{315} \;{\zeta_2}^3 
+ \frac{412}{5} \;{\zeta_2}^2 
\nonumber \\
&+ 80 \;{\zeta_2} {\zeta_3} 
-\frac{130}{3} \;{\zeta_2} 
+\frac{10336}{3} \;{\zeta_3}^2 - 460 \;{\zeta_3} 
+ 1328 \;{\zeta_5} - \frac{5599}{6}\Bigg) 
+ C_F^2 {n_f} \Bigg(\frac{272}{135} \;{\zeta_2}^2 
\nonumber \\
&- \frac{5504}{9} \;{\zeta_2} {\zeta_3}
+\frac{2632}{27} \;{\zeta_2} 
+ \frac{3512}{9} \;{\zeta_3} 
+ \frac{5536}{9} \;{\zeta_5} 
-\frac{421}{3}\Bigg)
+{C_F} n_{f,v} \Big( \frac{N^2 -4}{N} \Big)
\Bigg(-\frac{4}{5} \;{\zeta_2}^2
\nonumber \\
&+ 20 \;{\zeta_2} 
+ \frac{28}{3} \;{\zeta_3}
-\frac{160}{3} \;{\zeta_5} + 8 \Bigg) + {C_F} n_f^2 \
\Bigg(\frac{128}{27} \;{\zeta_2}^2 + \frac{2416}{81} \;{\zeta_2} 
- \frac{1264}{81} \;{\zeta_3} - \frac{7081}{243}\Bigg)\Bigg)
\nonumber \\
&+ C_A^2 {C_F} {{\cal D}_0} \Bigg(-\frac{2992}{15}\;{\zeta_2}^2-\frac{352}{3}\;{\zeta_2}{\zeta_3}+\frac{98224}{81}\;{\zeta_2}+\frac{40144}{27} \;{\zeta_3}-384\;{\zeta_5}-\frac{594058}{729}\Bigg)
\nonumber \\
&+C_A^2 {C_F} {{\cal D}_1}
\Bigg(\frac{704}{5} \;{\zeta_2}^2-\frac{12032 }{9}\;{\zeta_2}-704 \;{\zeta_3}
+\frac{124024}{81}\Bigg)
+C_A^2 {C_F} {{\cal D}_2}
\Bigg(\frac{704}{3} \;{\zeta_2}-\frac{28480}{27}\Bigg)
\nonumber \\
&+C_A^2 {C_F} {{\cal D}_3} \Bigg(\frac{7744}{27}\Bigg)
+{C_A} C_F^2 {{\cal D}_0}
\Bigg(\frac{1408}{3} \;{\zeta_2}^2-1472 \;{\zeta_2}\;{\zeta_3}-\frac{12416}{27} \;{\zeta_2}+\frac{26240}{9} \;{\zeta_3}
\nonumber\\
&+\frac{25856}{27}\Bigg)+{C_A} C_F^2 {{\cal D}_1}
\Bigg(\frac{3648}{5} \;{\zeta_2}^2-\frac{11648}{9} \;{\zeta_2}-5184\;{\zeta_3}-\frac{35572}{9}\Bigg)
\nonumber \\
&+{C_A} C_F^2 {{\cal D}_2}
\Bigg(\frac{11264}{3} \;{\zeta_2}+1344\;{\zeta_3}-\frac{4480}{9}\Bigg)+{C_A} C_F^2 {{\cal D}_3} \
\Bigg(\frac{17152}{9}-512\;{\zeta_2}\Bigg)
\nonumber \\
&+{C_A} C_F^2 {{\cal D}_4}\Bigg(-\frac{7040}{9}\Bigg)+{C_A} {C_F} {n_f} {{\cal D}_0}  \
\Bigg(\frac{736}{15} \;{\zeta_2}^2-\frac{29392}{81} \;{\zeta_2}-\frac{2480 }{9}\;{\zeta_3}+\frac{125252}{729}\Bigg)
\nonumber \\
&+{C_A} {C_F} {n_f} {{\cal D}_1}
\Bigg(384\;{\zeta_2}-\frac{32816}{81}\Bigg)+{C_A} {C_F} {n_f} {{\cal D}_2}  \Bigg(\frac{9248}{27}-\frac{128}{3} \;{\zeta_2}\Bigg)
\nonumber \\
&+{C_A} {C_F} {n_f} {{\cal D}_3} \Bigg( - \frac{2816}{27} \Bigg) 
+C_F^3 {{\cal D}_0} \Bigg(-6144 \;{\zeta_2}\;{\zeta_3}-4096\;{\zeta_3}+12288\;{\zeta_5}\Bigg)
\nonumber \\
&+C_F^3 {{\cal D}_1} \Bigg(-\frac{14208}{5} \;{\zeta_2}^2+2976\;{\zeta_2}-960\;{\zeta_3}+2044\Bigg)
+C_F^3 {{\cal D}_2} \Bigg(10240\;{\zeta_3} \Bigg)
\nonumber \\
&+C_F^3 {{\cal D}_3} \Bigg(-3072 \;{\zeta_2}-2048 \Bigg)+C_F^3 {{\cal D}_5} \Bigg( 512 \Bigg)+C_F^2 {n_f} {{\cal D}_0} \Bigg(-\frac{1472 }{15}\;{\zeta_2}^2+\frac{1952}{27}\;{\zeta_2}
\nonumber \\
&-\frac{5728}{9} \;{\zeta_3}-6\Bigg)+C_F^2 {n_f} {{\cal D}_1}
\Bigg(\frac{2048}{9} \;{\zeta_2}+1280\;{\zeta_3}+\frac{4288}{9}\Bigg)+C_F^2  {n_f} {{\cal D}_2}
\Bigg(\frac{544}{9} 
\nonumber \\
&-\frac{2048}{3} \;{\zeta_2}\Bigg) + C_F^2  {n_f} {{\cal D}_3} \Bigg(- \frac{2560}{9} \Bigg)+C_F^2  {n_f} {{\cal D}_4} \Bigg( \frac{1280}{9} \Bigg)+{C_F} n_f^2 {{\cal D}_0} \Bigg(\frac{640}{27}\;{\zeta_2}
\nonumber \\
&+\frac{320}{27} \;{\zeta_3}
-\frac{3712}{729}\Bigg)+{C_F} n_f^2 {{\cal D}_1}
\Bigg(\frac{1600}{81}-\frac{256}{9} \;{\zeta_2}\Bigg)- {C_F} n_f^2 {{\cal D}_2} \frac{640}{27}
\nonumber \\
&+ C_F n_f^2 {{\cal D}_3} \Bigg( \frac{256}{27} \Bigg)
\end{align}
where, $\zeta_i$ are the Riemann zeta functions, $C_F=(N^2-1)/2 N , C_A=N$ are the casimirs for $SU(N)$ gauge theory, $n_f$ is the number of active quark flavours and $n_{f,v}$ is the effective number of flavours resiling from some special class of diagrams at three loop~\cite{Gehrmann:2010ue}.
\section{Numerical Results}
\label{Results}
\begin{table}[h!]
\begin{center}
\begin{tabular}{| l | c | c | c | c | c | c | c  }
    \hline \hline
    ${\rm E_{CM}}$ & LO & ${\rm NLO_{SV}} $ & NLO & ${\rm NNLO_{SV}}$ & NNLO& ${\rm N^3LO_{SV}}$ \\
    \hline
    ~7 & 0.2415 & 0.2987 & 0.3183 & 0.3203 & 0.3257 & 0.3254 \\
    \hline
    ~8 & 0.2977 & 0.3667 & 0.3901 & 0.3932 & 0.3993 & 0.3991 \\
    \hline
    13 & 0.6120 & 0.7363 & 0.7788 & 0.7900 & 0.7975 &0.7970 \\
    \hline
    14 & 0.6801 & 0.8150 & 0.8604 & 0.8730 & 0.8808 & 0.8807 \\
    \hline \hline
 \end{tabular}
 \caption{DY like contributions (in pb) for different center of mass energies (TeV) at LHC with MSTW2008 PDFs. The factorization and renormalization scales are set to $\mu_F = \mu_R = Q$.}
 \label{DY}
 \end{center}
\end{table}
In what follows we present the numerical results for associated production of the
Higgs boson with vector boson at the LHC for the proton-proton center of mass energies of 
$7,8,13$ and $14$ TeV. The hadronic
cross sections are obtained by folding the respective LO, NLO and NNLO partonic cross 
sections with the parton distribution functions (PDFs) measured at the same order
in the perturbation theory and by using the corresponding strong coupling constant 
$\alpha_s(\mu_R)$.  For $\rm N^3LO$ threshold corrections, however, we use NNLO PDFs and 
the $\alpha_s(\mu_R)$ obtained from the $4-loop$ $\beta$~function. Unless mentioned 
otherwise, we use MSTW2008 PDFs for our results.  Except for the scale uncertainties, 
both the renormalization and the factorization scales are set to $\mu_R = \mu_F = Q$, 
where $Q^2=(p_V + p_H)^2$ is the invariant mass of the gauge boson and the Higgs boson.

For the numerical implementation of the $\rm N^3LO$ threshold corrections, we have included the
additional subroutines for the contributions coming from the $\delta(1-z)$ term and the logarithmic contributions 
${\cal D}_k$, in the code {\tt vh@nnlo} in a similar fashion as at the $2$-loop level.
This easily enables one to compute the $\rm N^3LO$ threshold corrections using the PDFs supplied
by {\tt LHAPDF} and the strong coupling constant as in the code {\tt vh@nnlo}.

First, we present the DY type contributions to the $ZH$ associated production up to 
${\rm N^3LO}$ in QCD for different LHC energies in table~\ref{DY}. Here NLO$_{\rm SV} =$ LO+$a_s \Delta_{q\overline{q}}^{\rm SV, (1)}$,
${\rm NNLO_{\rm SV}}$ = NLO + $a_s^2 \Delta_{q\overline{q}}^{\rm SV,(2)}$ and ${\rm N^3LO_{\rm SV}}$ = NNLO + $a_s^3 \Delta_{q\overline{q}}^{\rm SV,(3)}$. We observe 
that the soft plus virtual contributions make up to $75$\% of the exact QCD correction 
at NLO level while they are about $60$\% at NNLO level, showing the significant 
contribution of the large logarithms that arise in the threshold limit.
\begin{figure}[h!]
\centerline{ 
\includegraphics[width=8cm]{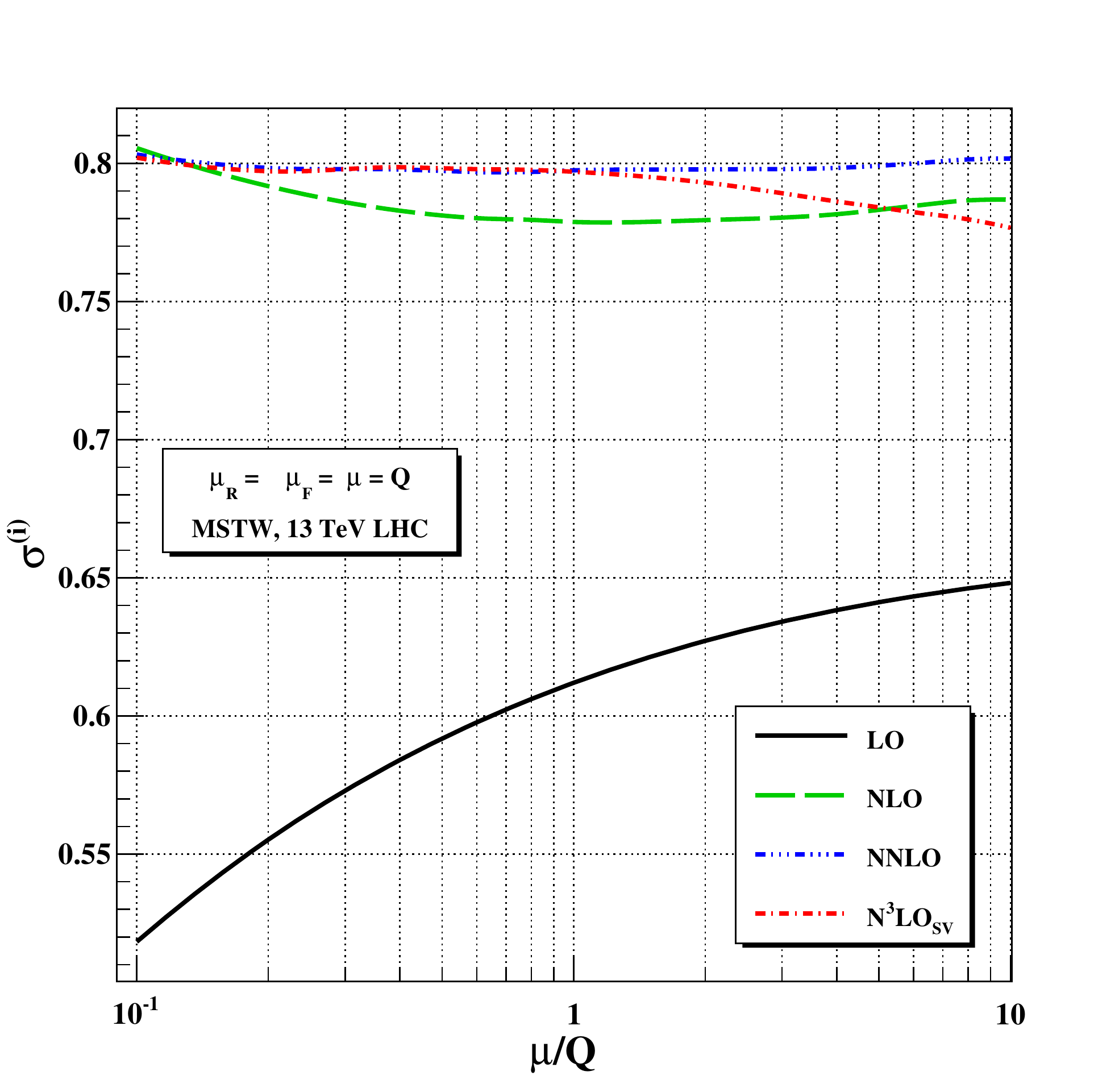}
}
\caption{\label{mu-var} Scale uncertainties of DY type cross sections for LHC13 by varying the 
factorization and renormalization scales in the range $0.1 < \mu/Q < 10.0$, where $\mu=\mu_F=\mu_R.$ 
}
\end{figure}
\begin{figure}[h!]
\centerline{ 
\includegraphics[width=8cm]{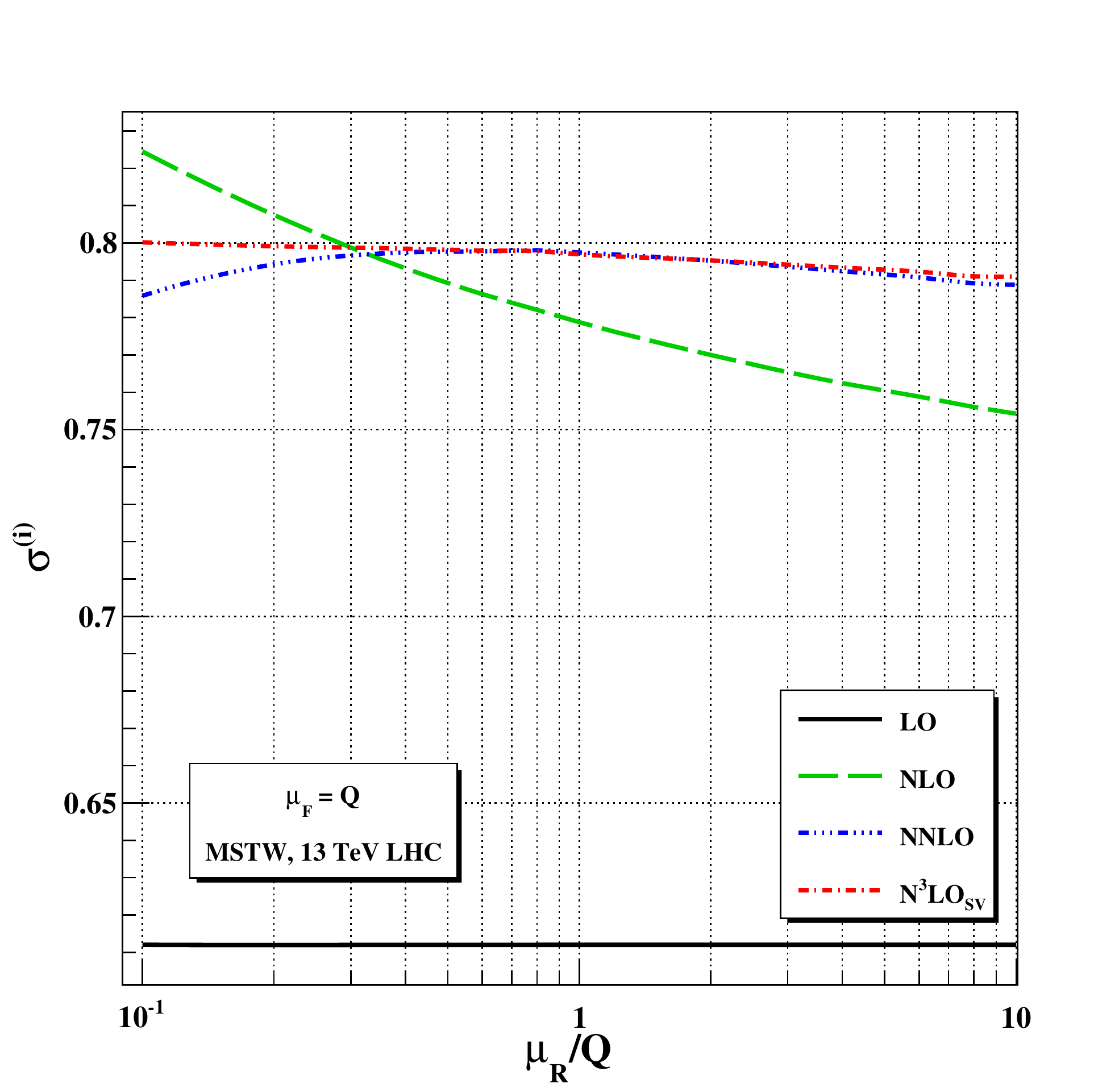}
\includegraphics[width=8cm]{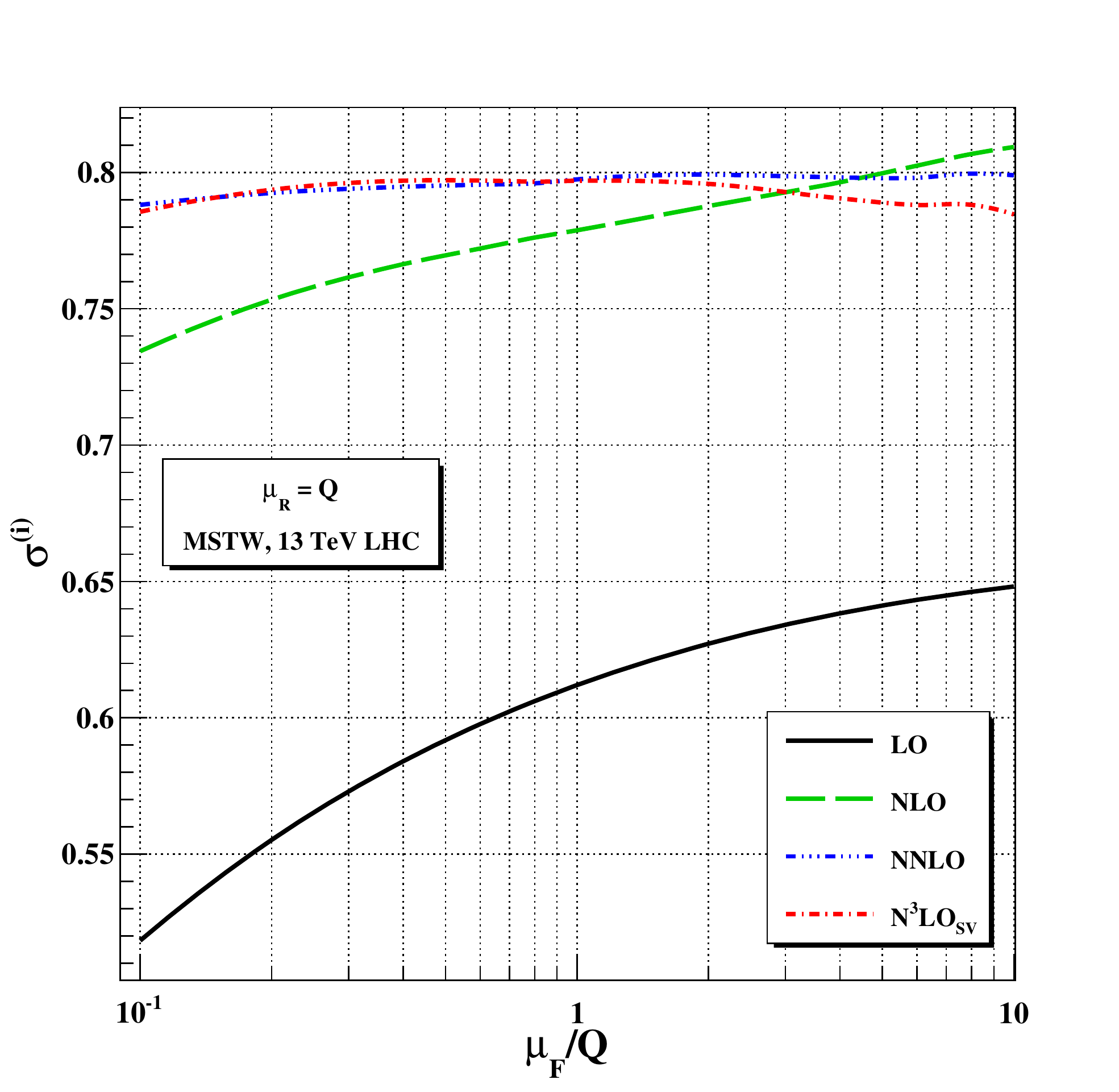}
}
\caption{\label{mf-mr-var} Scale uncertainties of DY type cross sections for LHC13.  In the left panel, we show
the renormalization scale uncertainty for $0.1 < \mu_R/Q < 10.0$ keeping $\mu_F=Q$ fixed.  In the right panel, we show
the factorization scale uncertainty for the similar range variation as $\mu_R$.
}
\end{figure}
The first 
and second order SV corrections are found to be positive and enhance the 
cross sections while the third order one is found to be negative for all different energies.
We also observe here that at $3$-loop level the $\delta(1-z)$ term can contribute as much as 
the ${\cal D}_{k}$ terms in magnitude. It can be noted that the impact of the QCD corrections 
increases with the decrease in the proton-proton collision energy.

\begin{table}
\begin{center}
\begin{tabular}{| l | c | c | c | c | c |}
    \hline \hline
    ${\rm E_{CM}}$ & LO & NLO & NNLO& ${\rm N^3LO_{SV}}$ \\
    \hline
    ~7 &  0.2292 & 0.3021 & 0.3230 & 0.3227 \\
    \hline
    ~8 & 0.2826 & 0.3702 & 0.3984 & 0.3982  \\
    \hline
    13 & 0.5797 & 0.7377 & 0.8146 & 0.8141  \\
    \hline
    14 & 0.6440 & 0.8148 & 0.9037 & 0.9035  \\
    \hline \hline
 \end{tabular}
 \caption{Total cross sections (in pb) for different center of mass energies (in TeV) at LHC.}
 \label{totalmstw}
 \end{center}
\end{table}
\begin{table}
\begin{center}
\begin{tabular}{| l | c | c | c | c | c | }
    \hline \hline
    PDFs & LO & NLO & NNLO& ${\rm N^3LO_{SV}}$ \\
    \hline
    MSTW2008 & 0.5797 & 0.7377 & 0.8146 & 0.8141 \\
    \hline
    ABM11 & -- & 0.7716 & 0.8308 & 0.8305  \\
    \hline
    NNPDF & 0.6199 & 0.7234 & 0.7997 & 0.7994  \\
    \hline
    CT10 & 0.6307 & 0.7312 & 0.8132 & 0.8128  \\
    \hline \hline
 \end{tabular}
 \caption{Total cross sections (in pb) for different PDFs at $13$ TeV LHC.}
 \label{totalpdf}
 \end{center}
\end{table}
Next, we study the scale uncertainties by varying the arbitrary factorization and renormalization
scales. In fig.\ref{mu-var}, we show the scale dependence of the DY like cross sections up to 
${\rm N^3LO_{SV}}$ by varying the scales in the range $0.1 < \mu/Q < 10.0$, where $\mu= \mu_R=\mu_F$.
The scale uncertainties are found to decrease with the order in the perturbation theory.  Here, at ${\rm N^3LO}$ only the soft plus virtual corrections are available. However, with the availability of the respective hard functions and the PDFs, the scale uncertainty is expected to improve further.

In the right panel of fig.\ref{mf-mr-var}, we show only the factorization scale dependence of the DY like
cross sections by varying $\mu_F$ in the range $0.1<\mu_F/Q < 10.0$ and keeping $\mu_R = Q$ fixed.  
The observations are similar to those found in fig.\ref{mu-var}. In the left panel of fig.\ref{mf-mr-var},
we show the renormalization scale dependence by varying it in the range $0.1 < \mu_R / Q < 10.0 $ and 
keeping $\mu_F=Q$ fixed.  Here the ${\rm N^3LO}$ scale uncertainties are found to be more stable than
the lower order results as expected.

Apart from the DY like contributions, there will also be other subprocess contributions 
such as $gg \to ZH$ via quark loops, 
$q \bar{q} \to Z H$ via {\it top}-loops at NNLO level.
Moreover, electroweak corrections for this process are already available and they do not depend on the QCD parameters.  For consistency, we include 
in our analysis all these contributions as in \cite{Brein:2003wg,Brein:2004ue,Brein:2011vx} and the corresponding third order result is given by
\begin{eqnarray}
\sigma^{\hbox{tot}}_{N^3LO} & = & \sigma^{DY}_{N^3LO} \left(1 + \delta_{EW} \right) + \sigma^{\hbox{gg}} + \sigma^{\hbox{top}}
\end{eqnarray}
%
In table~\ref{totalmstw}, we present the total cross sections up to $\rm N^3LO$ in QCD for different center of mass energies.
For LHC7 and LHC8, the gluon initiated subprocess contributions are about 5\% of DY type at NNLO 
while the EW corrections are
of the same size but with opposite sign.  Consequently, the total NNLO cross sections here are almost the same
as those of pure DY contributions.  However, for LHC13 and LHC14, the gluon initiates subprocess contributions
rises to about 9\% making the total cross sections larger than those of DY type.  In all these cases, the third
order QCD corrections are about $0.1\%$ but negative. 
Finally in table~\ref{totalpdf}, we present the total cross sections up to N$^3$LO for LHC13 for different parton 
distribution functions, namely, ABM11, CT10, NNPDFs and MSTW2008 PDFs.
\section{Conclusion}
\label{Conclude}
In this work we have computed the $\rm N^3LO$ QCD threshold corrections to 
the associated production of the Higgs with vector boson using
the inclusive third order DY corrections, which became available very recently.  With both the 
threshold logarithms ${\cal D}_k$ and the $\delta(1-z)$ term, these results are expected
to augment the previously available exact NNLO results for this process.  For the numerical computation, we have 
incorporated these corrections in the code {\tt vh@nnlo} to obtain the state of the art results.  
For our predictions, we have restricted ourselves only to  
the $ZH$ associated production and similar predictions can be made for $WH$ as well.
We have also estimated the theory uncertainties
from the factorization and renormalization scales and from the choice of the parton 
distribution functions.  While the hard part at the $\rm N^3LO$ level is yet to be computed, 
we believe that these results, providing the first predictions in this direction towards the computation of the full $\rm N^3LO$ for Higgs-Strahlung processes,
will be useful for the phenomenological studies related to Higgs Physics at LHC.

\section*{Acknowledgement}
M.C.K, M.K.M and VR would like to thank Stefano Frixione for suggesting this project and useful discussions. We also would like to thank Robert V. Harlander for his help with the code {\tt vh@nnlo}. M.K.M thanks IMSc for providing hospitality, where this work has been carried out. We thank T. Ahmed and N. Rana for discussions. The work of M.K.M has been partially supported by funding from Regional Center for Accelerator-based Particle Physics (RECAPP), Department of Atomic Energy, Govt. of India.

\bibliography{VHN3LO}
\bibliographystyle{utphysM}
\end{document}